\documentclass[reqno,11.5pt]{article}
\usepackage[normalem]{ulem}

\usepackage{booktabs} % To thicken table lines

\usepackage{amssymb}
\usepackage{tikz}
\usepackage{amsmath}
\usepackage{latexsym}
\usepackage{mathrsfs}
\usepackage{amsthm}
\usepackage{verbatim}
\usepackage{graphicx}
\usepackage{epstopdf}
\usepackage{epsfig}
\usepackage{color}
\usepackage{float}
\usepackage{titlesec}
\usepackage{bm}
\usepackage{amsfonts}
\usepackage{fancyhdr}
\usepackage{amscd}
\usepackage{cases}
\usepackage{amsmath,stackrel}

\usepackage{mathtools}

\usepackage{multirow}

\makeatletter
\@addtoreset{equation}{section}
\makeatother

\usepackage{geometry}
\usepackage{tikz}
\usetikzlibrary{er}

\usepackage[numbers]{natbib}
\usepackage[unicode=true,pdfusetitle,
bookmarks=ture,bookmarksnumbered=false,bookmarksopen=false,
breaklinks=false,pdfborder={0 0 1},backref=false,colorlinks=false]{hyperref}

\title{Constrained Linear Data-feature Mapping \\for Image Classification}

\author{Juncai He\footnotemark[1] \qquad Yuyan Chen\footnotemark[2] \qquad Lian Zhang\footnotemark[3] \qquad Jinchao Xu\footnotemark[4]}
\date{} % 去掉日期

\usepackage{algorithm, algorithmic}
\usepackage{lipsum}

\usepackage{tikz}   %commented by Ludmil, uncommented by Hongxuan
\usetikzlibrary{positioning}
\usepackage{amsmath}
\usepackage{amsfonts} 
\usepackage{amssymb}
\usepackage{bm}
\newtheorem{theorem}{Theorem}

\begin{document}
	% \nipsfinalcopy is no longer used
	
\maketitle 
\renewcommand{\thefootnote}{\fnsymbol{footnote}} %将脚注符号设置为fnsymbol类型，即特殊符号表示
\footnotetext[1]{Department of Mathematics, The Pennsylvania State University, University Park, PA 16802, USA (juh380@psu.edu).} %对应脚注[1]
\footnotetext[2]{School of Mathematical Sciences, Peking University, Beijing 100871, China (chenyuyan@pku.edu.cn).} %对应脚注[2]
\footnotetext[3]{Department of Mathematics, The Pennsylvania State University, University Park, PA 16802, USA (luz244@psu.edu).} %对应脚注[3]
\footnotetext[4]{Department of Mathematics, The Pennsylvania State University, University Park, PA 16802, USA (xu@math.psu.edu).} %对应脚注[4]

\maketitle
%\thispagestyle{empty}

%%%%%%%%% ABSTRACT
%\input{abstract}
\begin{abstract}	
In this paper, we propose a constrained linear data-feature mapping model 
as an interpretable mathematical model for image classification using convolutional
neural network (CNN) such as the ResNet. 
From this viewpoint, we establish the detailed connections in a technical level 
between the traditional iterative schemes for constrained linear system 
and the architecture for the basic blocks of ResNet. 
Under these connections, we propose some natural modifications of ResNet type
models which will have less parameters but still maintain almost the same accuracy
as these corresponding original models.
Some numerical experiments are shown to demonstrate the validity of 
this constrained learning data-feature mapping assumption. 
\end{abstract}

%%%%%%%%% BODY TEXT
%\input{introduction}
\section{Introduction}
This paper is devoted to providing some mathematical insight of deep
convolutional neural network models that have been successfully
applied in many machine learning and artificial intelligence areas
such as computer vision, natural language precessing and reinforcement
learning~\cite{lecun2015deep}.
Important examples of CNN include the LeNet-5 model of LeCun et al. in
1998~\cite{lecun1998gradient}, the AlexNet of Hinton et el in 2012
\cite{krizhevsky2012imagenet}, Residual Network (ResNet) of K. He et
al in 2015~\cite{he2016deep} and Pre-act ResNet in
2016~\cite{he2016identity}, and other variants of CNN in
\cite{simonyan2014very, szegedy2015going, huang2017densely}.  
Among these different CNNs, ResNet and pre-act ResNet models are of
special theoretical and practical interests.  It has been an active
research topic on theoretical understanding or explanation of why and
how ResNet work well, and how to design better residual type
architectures based on certain empirical observations and formal
interpretation, see \cite{zagoruyko2016wide, larsson2016fractalnet,
	gomez2017reversible, xie2017aggregated, zhang2017polynet,
	szegedy2017inception, huang2017densely}.  For example, a dynamical
system viewpoint  was discussed in~\cite{lu2018beyond,
	chen2018neural} to explain the rational for skip connections in
ResNets.

In this paper, we propose a generic mathematical model behind the
residual blocks in ResNet to understand how ResNet model works.
The core of our model is the following assumption:  there is a data-feature 
mapping
\begin{equation}\label{Auf}
A\ast u = f,
\end{equation}
where $f$ is the data such as images we see and 
$u$ represents a feature tensor such that
\begin{equation}
\label{positive-u}
u\ge 0.  
\end{equation}
Feature extraction is then viewed as an iterative procedure
(c.f. \cite{xu1992iterative}) to solve \eqref{Auf}:
\begin{equation}
\label{iterative0}
u^{i} = u^{i-1} + B^i\ast (f - A\ast u^{i-1}), \quad i=1:\nu.
\end{equation}
Using,  for example,  the special activation function $\sigma(x)={\rm
	ReLU}(x) := \max\{0,x\}$, the above iterative process can be naturally modified to
preserve the constraint
\eqref{positive-u}: 
\begin{equation}
\label{iterative}
u^{i} = u^{i-1} + \sigma \circ B^i\ast \sigma  (f -  A
\ast u^{i-1}), \quad i=1:\nu.
\end{equation}
Introducing the residual
\begin{equation}
\label{residual}
r^i=f-A\ast u^i,   
\end{equation}
the iterative process \eqref{iterative} can be written as
\begin{equation}
\label{modified-resnet}
r^{i} = r^{i-1} - A\ast \sigma \circ B^i\ast\sigma(r^{i-1}), \quad i=1:\nu.
\end{equation}
This above process represent one major modified pre-act ResNet to be studied
in this paper and it can be directly compared with the following process
representing a core component of pre-act ResNet \cite{he2016identity}:
\begin{equation}
\label{resnet1}
r^{i} = r^{i-1} - A^i\ast \sigma \circ B^i\ast\sigma(r^{i-1}), \quad i=1:\nu.  
\end{equation}
At least two observations can be made by comparing 
\eqref{modified-resnet} with \eqref{resnet1}: (1)
The $A^i$ in the pre-act ResNet \eqref{resnet1} depends on $i$,
whereas the $A$ in the modified pre-act ResNet \eqref{modified-resnet} does not
depend on $i$. (2) The classic ResNet such as \eqref{resnet1} can be related to
iterative methods for solving systems of equations. 
These two observations represent the key ideas of this
paper. 

Furthermore, by involving the multigrid~\cite{xu1992iterative,hackbusch2013multi} idea 
about how to restrict the residuals, we have a natural explanation for pooling 
operation in pre-act ResNet. This helps us  establish a complete connection 
between pre-act ResNet and MgNet which is proposed in~\cite{he2019mgnet}.
We will provide some numerical evidences to demonstrate that our constrained
linear model \eqref{Auf} and \eqref{positive-u} with the nonlinear iterative solvers
\eqref{iterative} or \eqref{modified-resnet} provide a good interpretation and
improvement of ResNet.
Our main contributions can be summarized as: 
(1) Propose and develop the constrained linear data-feature mapping
assumption as an interpretable model for ResNet models. 
(2) Propose some natural modifications of ResNet type models based 
on this interpretation and demonstrate their efficiency on some standard datasets.
(3) Provide both theoretical and numerical validation of the special 
schemes of linear data-feature mapping and nonlinear solver.

\subsection{Related works}
The data-feature mapping is first proposed in~\cite{he2019mgnet}, 
which establish the connection between ResNet type CNNs 
and multigrid methods. Under this assumption, \cite{he2019mgnet}
proposes a new architecture, known  as MgNet, by applying the iterative scheme to
a constrained linear model \eqref{Auf} and \eqref{positive-u}.
Before MgNet, ideas and techniques from multigrid 
methods have been used for the development of efficient
CNNs.  The authors in ResNet~\cite{he2016deep} first took
the multigrid methods as the evidence to support their so-called residual representation interpretation 
for ResNet.
Besides this, \cite{ke2016multigrid, haber2017learning, zhang2019scan} adopt the multi-resolution
ideas to enhance the performance of their networks. 
Furthermore, a CNN model whose  structure is similar to the V-cycle multigrid is 
proposed to deal with volumetric medical image segmentation and 
biomedical image segmentation in~\cite{ronneberger2015u, milletari2016v}.  
There are also some literature about applying deep learning
techniques into multigrid or numerical PDEs such as~\cite{katrutsa2017deep,hsieh2018learning}.

Considering the connections of CNN models and methods in computational mathematics,
researchers also propose the dynamic system or optimization perspective~
\cite{haber2017learning,e2017a, chang2017multi, lu2018beyond, chen2018neural}.
One key reason why people propose the viewpoint of dynamic systems
is that the iterative scheme $x^i = x^{i-1} + f(x^{i-1})$ in pre-act ResNet resemble the
forward Euler scheme in numerical dynamic systems. 
Following this idea, \cite{sonoda2017double, li2017a} interpreted the date flow in ResNet as the solution of transport equation following the characteristic line.
Furthermore, \cite{lu2018beyond} interpreted some different CNN models with residual block  like 
PloyNet~\cite{zhang2017polynet}, FractalNet~\cite{larsson2016fractalnet} and RevNet~\cite{gomez2017reversible}
as some special discretization schemes for ordinary differential equations (ODEs).
Ignoring the specific discretization methods, \cite{chen2018neural} proposes a family of CNNs based on any black box solvers 
for ODEs. 
Some CNN architectures are further designed based on the iterative schemes of optimization algorithms
like~\cite{gregor2010learning, sun2016deep, li2018optimization}. 
The aforementioned works share a common philosophy that many optimization algorithms
can be considered as certain discretization schemes for some special ODEs~\cite{helmke2012optimization}.

\section{ResNet and Pre-act ResNet with Mathematical Formula}
\begin{figure}[H]
	\centering
	\includegraphics[width=0.8\linewidth]{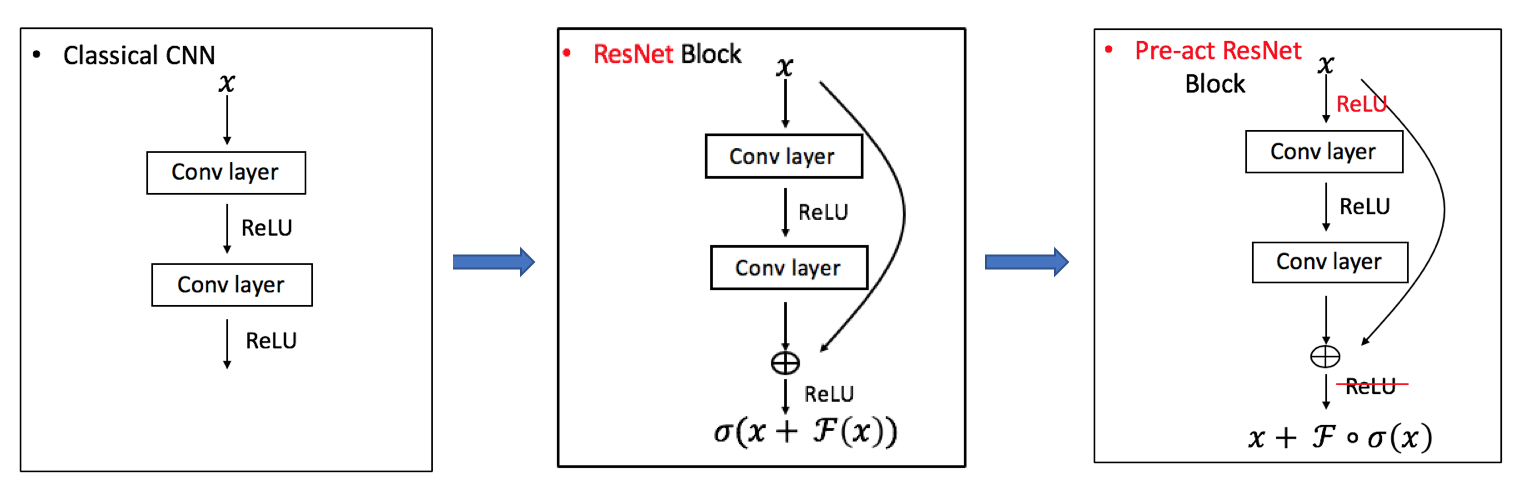}
	\caption{Comparison of classical CNN, ResNet and pre-act ResNet}
	\label{fig:resent}
\end{figure}

Let us first use the Figure~\ref{fig:resent} to demonstrate the connection and 
difference between classical CNN, ResNet~\cite{he2016deep} and pre-act ResNet~\cite{he2016identity}.
Here we  use the notation that $\sigma(x) = {\rm ReLU}(x) := \max\{0,x \}$ as the standard ReLU activation function.
For ResNet and pre-act ResNet with basic block, $\mathcal F(x) = A \ast \sigma \circ B \ast  x$ where $A$ and $B$
are $3\times3$ convolutions with multichannel, zero padding and stride one and ``$\circ$'' means composition.
Our goal here is to investigate the interpretable mathematical model behind these models. 
To do that, let us first try to write these CNN models with specific mathematical formulas.
\paragraph{Pre-act ResNet}
Different from the general researches by using diagram to identify
these CNN architectures, we use some exact mathematical formula 
to write CNN models.
One main component in the pre-act ResNet~\cite{he2016identity} without the last fully connected and soft-max layers, 
it can be written as in Algorithm \ref{alg:presnet}.
\begin{algorithm}[H]
	\footnotesize
	\caption{$ h = \text{pre-act ResNet}(f; J,\nu_1, \cdots, \nu_J)$}
	\label{alg:presnet}
	\begin{algorithmic}[1]
		\STATE Initialization:  $r^{1,0} = f_{\rm in}(f)$.
		%		\STATE Initialization $u^{1,0}$
		\FOR{$\ell = 1:J$}
		\FOR{$i = 1:\nu_\ell$}
		\STATE Basic Block:
		\begin{equation}\label{ori-ResNet}
		r^{\ell,i} = r^{\ell, i-1} + A^{\ell,i} \ast  \sigma \circ B^{\ell,i}\ast   \sigma (r^{\ell,i-1}).
		\end{equation}
		\ENDFOR
		%		\STATE Note: $ u^\ell= u^{\ell,\nu_\ell} $
		\STATE Pooling(Restriction):
		\begin{equation}
		\label{ori-ResNet0}
		r^{\ell+1,0} = R_\ell^{\ell+1} \ast_2  r^{\ell, \nu_\ell} + A^{\ell+1,0} \circ \sigma \circ B^{\ell+1,0} \ast_2  \sigma (r^{\ell, \nu_\ell} ).
		\end{equation}
		\ENDFOR
		\STATE Final average pooling layer:
		$h =  R_{\rm ave}( r^{L,\nu_\ell})$.
	\end{algorithmic}
\end{algorithm}

Here $f_{\rm in}(\cdot)$ may depend on different data set and problems 
such as $f_{\rm in}(f) = \sigma \circ \theta^0 \ast f $ for CIFAR~\cite{krizhevsky2009learning} and
$f_{\rm in}(f) = R_{\rm max}\circ \sigma \circ \theta^0 \ast  f$ for ImageNet~\cite{deng2009imagenet} as in~\cite{he2016identity}.
In addition $r^{\ell,i} =   r^{\ell, i-1} +  A^{\ell,i} \ast  \sigma \circ B^{\ell,i} \ast \sigma (r^{i-1})$ is often called the basic ResNet block.
Here, $A^{\ell,i}$ with $i\ge0$ and $B^{\ell,i}$ with $i\ge1$ are general $3\times3$ convolutions with zero padding and stride 1.
In pooling block, $\ast _2$ means convolution with stride 2 and $B^{\ell,0}$ is taken as the $3\times3$ kernel with same output channel dimension of $R_\ell^{\ell+1}$
which is taken as $1\times1$ kernel and called as projection operator in \cite{he2016identity}. 
During two consecutive pooling blocks, index $\ell$ means the fixed resolution or we $\ell$-th grid level as in multigrid methods.
Finally, $R_{\rm ave}$ ($R_{\rm max}$) means average (max) pooling with different strides which is also dependent on datasets and problems.

\paragraph{ResNet}
The original ResNet, developed earlier in~\cite{he2016deep}, shares
almost the same scheme  with pre-act ResNet but with a different order of convolution 
and activation function.
For ResNet model, these basic block and pooling are defined by:
\begin{align}
r^{\ell,i} &= \sigma \left( r^{\ell, i-1} +  A^{\ell,i} \ast  \sigma \circ B^{\ell,i} \ast r^{\ell,i-1}\right), \label{eq:pre-actResNet1}\\
r^{\ell+1,0} &=  \sigma \left( R_\ell^{\ell+1} \ast _2 r^{\ell, \nu_\ell} + A^{\ell+1,0} \ast  \sigma \circ B^{\ell+1,0} \ast _2r^{\ell, \nu_\ell}  \right) , \label{eq:pre-actResNet2}
\end{align}
for $i = 1:\nu_\ell$.

\section{Constrained Linear Data-feature Mapping}
In this section, we will establish a new understanding of pre-act ResNet  
by involving the idea that the pre-act ResNet block is an iterative scheme for solving some 
hidden model in each grid.
We adopt this assumption into these ResNet type models 
and get some modified models with a special parameter sharing scheme.

\subsection{Constrained linear data-feature mapping and iterative methods}
%In this subsection, we will propose the constrained linear data-feature mapping
%behind these ResNet type CNNs as an interpretable model to understand
%what the iterative scheme in \eqref{ori-ResNet}, \eqref{eq:pre-actResNet1}
%are doing.
%Then we will unified these models based on this data-feature mapping.
%Because of the linearity of $A$, the above feature extraction process is 
%the so-called single step residual correction smoother in numerical
%linear algebra and multigrid methods~\cite{xu1992iterative,golub2012matrix,hackbusch2013multi}.
%The above iterative scheme \eqref{BAmapping}
%can be interpreted as both the feature extraction step in ResNet type models
%and the smoothing step in multigrid method.

The main point here is the introduction of 
the so-called data and feature space for CNN, which is analogous to the 
function space and its duality in the theory of multigrid 
methods~\cite{xu2017algebraic}. 
Namely, following~\cite{he2019mgnet} we introduce 
the next data-feature mapping model in every grid level follows:
\begin{equation}\label{eq:fmapping}
A^{\ell} \ast u^\ell = f^{\ell},
\end{equation}
where $f^\ell$ and $u^\ell$ belong to the data and feature space at $\ell$-th grid. 
We now make the following two important observations for this data-feature mapping:
\begin{itemize}
	\item The mapping in \eqref{eq:fmapping} is linear, more specifically it is just a convolution with multichannel, zero
	padding and stride one as in pre-act ResNet.
	\item In each level, namely between two consecutive pooling, there is only one
	data-feature mapping, or we say that $A^\ell$ only depends on $\ell$, but not on number of layers.
\end{itemize}
We note that this the assumption that these linear data-feature mapping
depend only on the grid level $\ell$ is motivated from a basic property of 
multigrid methods~\cite{xu1992iterative, hackbusch2013multi, xu2017algebraic}.

Besides \eqref{eq:fmapping}, we introducing an important constrained condition
in feature space that
\begin{equation}\label{eq:positive}
u^{\ell,i} \ge 0.
\end{equation}
The rationality of this constraint in feature space can be interpreted as follows.
First of all, from the real neural system, the real neurons will only be
active if the electric signal is greater than certain thresholding value. 
Namely, we can think that human brains can only see features 
with certain threshold.
On the other hand, the ``shift'' invariant property of feature space in CNNs,
namely, $u+a$ will not change the classification results. This means that $u+a$ should
have the same classification result with $u$. That is to say, we may assume that
$u \ge 0$ to reduce some redundancy of $u$.

Based on these assumptions above, what we need to do next is to
solve the data-feature mapping equation in \eqref{eq:fmapping}.
We will adopt some classical iterative methods~\cite{xu1992iterative} in scientific computing
to solve the system \eqref{eq:fmapping} and obtain that 
\begin{equation}\label{BAmapping}
u^{\ell,i} = u^{\ell,i-1} + B^{\ell,i} \ast (f^{\ell} - A^{\ell}\ast u^{\ell,i-1}),~ i = 1:\nu_\ell,
\end{equation}
where $u^{\ell} \approx u^{\ell,\nu_\ell}$. 
For more details about iterative methods
in numerical analysis, we refer to~\cite{xu1992iterative, hackbusch1994iterative, golub2012matrix}.
To preserve \eqref{eq:positive}, we naturally use the ReLU activation function $\sigma$
to modify \eqref{BAmapping} as follows
\begin{equation}
\label{eq:uBfAu}
u^{\ell,i} = u^{\ell,i-1} + \sigma \circ B^{\ell,i}\ast \sigma  (f^\ell -  A^\ell
\ast u^{\ell,i-1}), \quad i=1:\nu_\ell.
\end{equation}

Because of the linearity of convolution in data-feature mapping,
if we consider the residual $r^{\ell,j} = f^{\ell} - A^{\ell}\ast u^{\ell,j}$, 
\eqref{eq:uBfAu} leads to the next iterative forms for residuals
\begin{equation}\label{eq:pre-actResNet_residual}
r^{\ell, i} = r^{\ell,i-1} - A^\ell \ast \sigma \circ B^{\ell,i}\ast \sigma(r^{\ell,i-1}).
\end{equation}
This is the same as \eqref{eq:pre-actResNet1} under the constraint $A^{\ell,i} = A^{\ell}$ in pre-act ResNet.

%The next theorem shows the connections of iterative schemes between 
%$u^{\ell,i}$ in feature space and $r^{\ell,i}$ in residual form.
We summarize the above derivation in the following simple theorem.
\begin{theorem}\label{thm:1} Under the assumption that there is only
	one linear data-feature mapping in each grid $\ell$, i.e. $A^{\ell,i} = A^{\ell}$, 
	the iterative form in feature space as in \eqref{BAmapping} is equivalent to 
	\eqref{eq:pre-actResNet_residual} if $A^\ell$ is invertible where $r^{\ell,i} = f^\ell - A^{\ell}\ast u^{\ell,i}$.
\end{theorem}
%At last, we want to mention an unexpected discover that 
%$r^{\ell,i}$ is the residual in each steps for solving $A^{\ell}\ast u^\ell = f^\ell$, 
%which means that these ResNet or pre-act ResNet CNN models are really 
%dealing with ``residuals'' in some sense.

\subsection{Modified pre-act ResNet and ResNet}
\begin{figure}[H]
	\centering
	\includegraphics[width=0.38\linewidth]{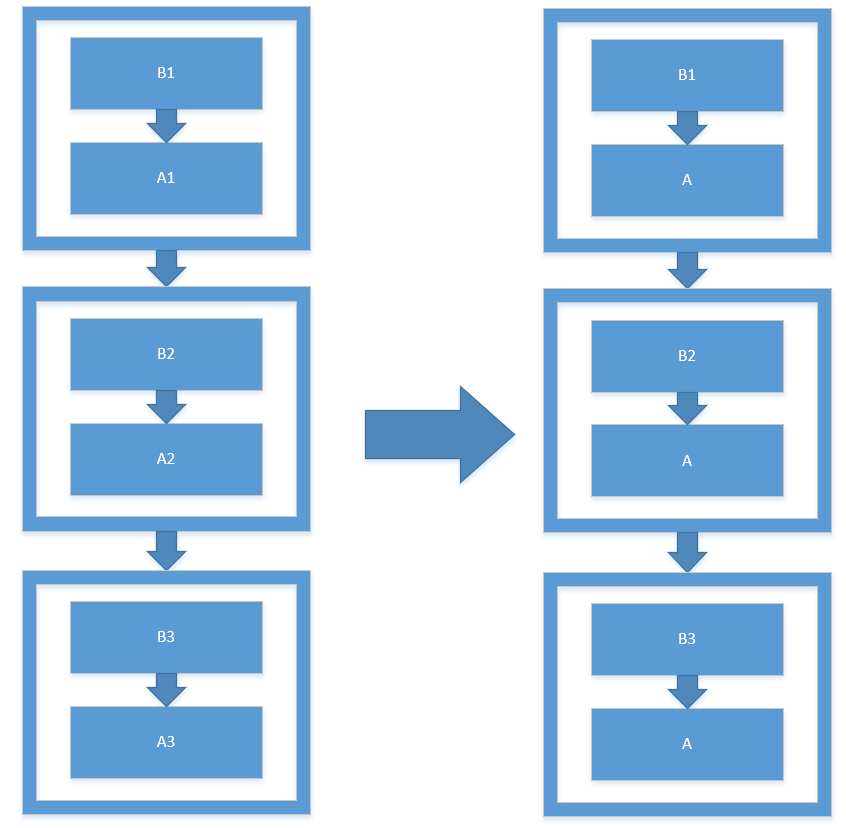}
	\caption{Diagram of modified (pre-act) ResNet basic blocks.}
	\label{fig.shareing}
\end{figure}

In this subsection we will propose some modified ResNet and pre-act ResNet 
models based on the assumption of the constrained linear 
data-feature mapping behind these models.
While the scheme in \eqref{eq:pre-actResNet_residual} is closely
related to the original pre-act ResNet,there is a major difference with an extra
constraint that $A^{\ell,i} = A^{\ell}$. 
As a result, we obtain these next modified pre-act ResNet  first as
\paragraph{Modified Pre-act ResNet (Pre-act ResNet-$A^\ell$-$B^{\ell,i}$)}
\begin{equation}\label{eq:rpre-actResNet}
r^{\ell,i} =r^{\ell, i-1} + A^{\ell} \ast \sigma \circ B^{\ell,i} \ast \sigma (r^{\ell,i-1}).	
\end{equation}
Here, we make a small modification of the sign before $A^\ell$ in formula since the linearity of convolution.
As we discussed before, the modified pre-act ResNet model
is derived from the constrained linear data-feature mapping 
by using a special iterative scheme.
Although we cannot get these connections in ResNet directly, formally we can just make
the modification from $A^{\ell,i}$ to $A^{\ell}$ into \eqref{ori-ResNet}
to obtain the corresponding modified ResNet models as follows,
%\begin{description}
\paragraph{Modified ResNet (ResNet-$A^\ell$-$B^{\ell,i}$)}
\begin{equation}\label{eq:rResNet}
r^{\ell,i} = \sigma \left( r^{\ell, i-1} +  A^{\ell} \ast \sigma \circ B^{\ell,i} \ast r^{\ell, i-1}\right).	
\end{equation}
A unified but simple diagram ignoring the activation functions 
for these modified pre-act ResNet and  ResNet with this structure can be shown as in Figure~\ref{fig.shareing}.

\section{Linear versus nonlinear data-feature mapping}
In this section, we will try to investigate the rationality for the constrained linear
data-feature mapping. We will show that linear data-feature mapping
model is adequate by comparing with some special nonlinear models
on the  data-feature mapping. 
%In this section, we will firstly propose a general data-feature mapping model
%and state its iterative process in feature space as an algorithm.
%Then we will find that the constrained linear data-feature mapping
%can be written as a special choice of some operators in that model.
%At last, we will design a numerical experiment to show the reasonableness of
%the special setting as discussed before.

\subsection{Nonlinear data-feature mapping and iterative methods}
One of the most important assumptions above is that 
the data-feature mapping \eqref{eq:fmapping} could be a linear model and
there should be only one model in each grid. 
To demonstrate that this linear model is adequate for image classification,
we compare it with the following nonlinear data-feature mapping:
\begin{equation}\label{eq:gfmapping}
\mathcal A^{\ell}(u^\ell) = f^{\ell},
\end{equation}
where $\mathcal A^{\ell}$ can be chosen for some special nonlinear forms, such as 
$A^\ell \ast \sigma$, $\sigma \circ A^\ell \ast$, and $\sigma \circ A^\ell \ast \sigma$.
Then we have the next iterative feature extraction scheme:
\begin{equation}\label{gBAmapping}
u^{\ell,i} = u^{\ell,i-1} + \mathcal B^{\ell,i} (f^{\ell} - \mathcal A^{\ell}(u^{\ell,i-1})),~ i = 1:\nu_\ell,
\end{equation}
where $\mathcal B^{\ell,i}$ can also take some special nonlinear forms.
Here we note that, because of the nonlinearity of $\mathcal A^{\ell}$, we cannot got the 
iterative scheme about  the residuals for \eqref{gBAmapping}. 
Namely, we can only do iteration in the feature space.
Thus, we propose the next feature based ResNet (FB-ResNet) in Algorithm \ref{alg:FB-ResNet}, which
consists of the iteration of features as in \eqref{gBAmapping} and a special pooling~\eqref{interpolation}.
\begin{algorithm}[H]
	\footnotesize
	\caption{$u^{J,\ell_J}=\text{ FB-ResNet}(f; J,\nu_1, \cdots, \nu_J)$}
	\label{alg:FB-ResNet}
	\begin{algorithmic}[1]
		\STATE Initialization:  $f^1 = f_{\rm in}(f)$, $u^{1,0}=0$
		%		\STATE Initialization $u^{1,0}$
		\FOR{$\ell = 1:J$}
		\FOR{$i = 1:\nu_\ell$}
		\STATE Feature extraction (smoothing):
		\begin{equation}\label{mgnet}
		u^{\ell,i} = u^{\ell,i-1} + \mathcal B^{\ell,i}  \left({f^\ell -  \mathcal A^{\ell} (u^{\ell,i-1})}\right).
		\end{equation}
		\ENDFOR
		%		\STATE Note: $ u^\ell= u^{\ell,\nu_\ell} $
		\STATE Pooling (interpolation and restriction):
		\begin{equation}
		\label{interpolation}
		\begin{aligned}
		u^{\ell+1,0} &= \Pi_\ell^{\ell+1} \ast_2  u^{\ell, \nu_\ell} \\
		%		\end{equation}
		%		\begin{equation}
		%		\label{restrict-f}
		f^{\ell+1} &= R^{\ell+1}_\ell \ast_2 (f^\ell - \mathcal A^\ell(u^{\ell, \nu_\ell}) + \mathcal A^{\ell+1} (u^{\ell+1,0}).
		\end{aligned}
		\end{equation}
		\ENDFOR
	\end{algorithmic}
\end{algorithm}
Here \eqref{interpolation} are understood as different interpolation
and restriction operators because of the difference of the feature and data space. 
However, in really implementation there are all implemented by $3\times3$ convolution with stride 2
as $\Pi_\ell^{\ell+1} \ast_2$ and $R^{\ell+1}_\ell \ast_2$.
We note that the idea to use the feature as the iterative unit is also proposed  in MgNet~\cite{he2019mgnet}.
More detailed discuss about the relation and comparison between typical pooling operation such as~\eqref{ori-ResNet0}
and restriction in \eqref{interpolation} could be found in Section 1 in the {\bf supplementary materials}.

%Actually, the special restriction (pooling) structure in \eqref{interpolation}
%first appear also in~\cite{he2019mgnet}.

\subsection{Numerical comparisons}
To investigate the optimality of the linear assumption of $\mathcal A^{\ell}$,
us first assume that we still keep the linearity assumption about $\mathcal A^{\ell}$ 
with the iterative method \eqref{gBAmapping}.
Then we can have the next iterative scheme for residuals $r^{\ell,i} = f^\ell - \mathcal A^{\ell}(u^{\ell,i})$ as:
\begin{equation}\label{eq:gresidual-iter1}
r^{\ell, i} = r^{\ell,i-1} - \mathcal A^\ell \mathcal B^{\ell,i}(r^{\ell,i-1}).
\end{equation}
If we take the following specific settings: $ \mathcal A^\ell(u) = A^\ell \ast u$ 
and $\mathcal B^{\ell,i}(r) = \sigma \circ B^{\ell,i} \ast \sigma(r)$.
The iterative scheme for residuals will becomes
\begin{equation}
r^{\ell, i} = r^{\ell,i-1} - A^\ell \ast \sigma \circ  B^{\ell,i}\ast \sigma(r^{\ell,i-1}),
\end{equation}
which is exact the modified pre-act ResNet scheme as we discussed before.

Thus, we try some numerical experiments with ``symmetric'' forms for different linear or nonlinear forms for both
$\mathcal A^\ell$ and $ \mathcal B^{\ell,i}$ as:
\begin{equation}
K\ast, ~K \ast \sigma,~ \sigma \circ K\ast, ~\text{and}~\sigma \circ K \ast \sigma,
\end{equation}
where $K$ is a $3\times3$ convolution kernel with multichannel, zero padding
and stride one.
These models can also be understood with the similar idea in pre-act ResNet,
which are obtained by moving these activation functions and convolutions around in ResNet.
In some sense, this is another important reason that why we propose the FB-ResNet as it is in feature space.
Otherwise,  it will be too close to the method in developing pre-act ResNet as in~\cite{he2016identity}
if we still take its iterative scheme for residual form.
The next table shows the numerical results with different combinations
of linear or nonlinear schemes for $\mathcal A^\ell$ and $\mathcal B^{\ell,i}$.
\begin{table}[H]
	\caption{TOP-1 accuracy of models from Algorithm~\ref{alg:FB-ResNet} with different linear or non-linear schemes of $\mathcal A$ and $\mathcal B$ on CIFAR10.}
	\label{tab:compare-AB}
	\begin{center}
		\begin{tabular}{lc}
			\toprule
			Schemes of $\mathcal A$ and $\mathcal B^{\ell,i}$   & Accuracy	\\
			\midrule
			$\mathcal A^{\ell} = A^{\ell} \ast$,~ $\mathcal B^{\ell,i} = B^{\ell,i}\ast $  	&70.96 \\
			%			\hline		%1
			$\mathcal A^{\ell} = A^{\ell} \ast$, ~$\mathcal B^{\ell,i} =  \sigma \circ B^{\ell,i} \ast$     &92.82 \\
			%			\hline		%6
			$\mathcal A^{\ell} = A \ast$, ~$\mathcal B^{\ell,i} = B^{\ell,i} \ast \sigma$     &93.01 \\
			%			\hline		%5
			{$\mathcal A^{\ell} = A^{\ell} \ast$, ~$\mathcal B^{\ell,i} = \sigma \circ B^{\ell,i} \ast \sigma $}     &{\bf 93.49} \\
			%			\hline
			$\mathcal A^{\ell} = A^{\ell} \ast \sigma $, ~$\mathcal B^{\ell,i}$ = $B^{\ell,i} \ast$     & 92.64 \\
			%			\hline
			$\mathcal A^{\ell} = A^{\ell} \ast \sigma $, ~$\mathcal B^{\ell,i} =  \sigma \circ B^{\ell,i} \ast$     & 92.54 \\
			%			\hline
			$\mathcal A^{\ell} = A^{\ell} \ast \sigma $, ~$\mathcal B^{\ell,i} = B^{\ell,i} \ast \sigma $   & 93.46 \\
			%			\hline
			$\mathcal A^{\ell} = A^{\ell} \ast \sigma $, ~$\mathcal B^{\ell,i} = \sigma \circ B^{\ell,i} \ast \sigma $     & 93.15 \\
			%			\hline
			$\mathcal A^{\ell} = \sigma \circ A^{\ell} \ast$, ~$\mathcal B^{\ell,i}$ = $B^{\ell,i} \ast$     &91.91 \\
			%			\hline
			$\mathcal A^{\ell} = \sigma \circ A^{\ell} \ast$, ~$\mathcal B^{\ell,i} =  \sigma \circ B^{\ell,i} \ast$    &92.14 \\
			%			\hline	%1
			$\mathcal A^{\ell} = \sigma \circ A^{\ell} \ast$, ~$\mathcal B^{\ell,i} = B^{\ell,i} \ast \sigma $ &93.37 \\
			%			\hline	%6
			$\mathcal A^{\ell} = \sigma \circ A^{\ell} \ast$, ~$\mathcal B^{\ell,i} = \sigma \circ B^{\ell,i} \ast \sigma $ 	&93.17 \\
			%			\hline	
			$\mathcal A^{\ell} = \sigma \circ A^{\ell} \ast \sigma$, ~$\mathcal B^{\ell,i}$ = $B^{\ell,i} \ast$   &92.70\\
			%			\hline
			$\mathcal A^{\ell} = \sigma \circ A^{\ell} \ast \sigma$, ~$\mathcal B^{\ell,i} =  \sigma \circ B^{\ell,i} \ast$   &93.23\\
			%			\hline
			$\mathcal A^{\ell} = \sigma \circ A^{\ell} \ast \sigma$, ~$\mathcal B^{\ell,i} = B^{\ell,i} \ast \sigma $   &93.37\\
			%			\hline 
			$\mathcal A^{\ell} = \sigma \circ A^{\ell} \ast \sigma$, ~$\mathcal B^{\ell,i} = \sigma \circ B^{\ell,i} \ast \sigma $   &93.40\\
			\bottomrule
		\end{tabular}
	\end{center}
\end{table}

From the results in Table \ref{tab:compare-AB}, we show that the
original assumption about the linearity of $\mathcal A^\ell$ and the special non-linear form of $\mathcal B^{\ell,i}$ 
is the most rational and accurate scheme which is
also consistent with the theoretical concern and numerical results as in this paper.

\section{Experiments}
Our numerical experiments indicate that
fixing the linear data-feature mapping in each grids only bring 
little negative or sometimes good effects than the standard ResNet and pre-act ResNet,
which demonstrate the rational of the constrained data-feature mapping model. 

\subsection{Datasets, models and training details}
\paragraph{Datasets}
We evaluate our methods on the following widely used datasets: 
MNIST~\cite{lecun1998gradient} , CIFAR datasets~\cite{krizhevsky2009learning}(CIFAR10, CIFAR100) and ImageNet~\cite{deng2009imagenet}. 
%I suggest to put some introduction materials about MNIST and CIFAR based on the style of DenseNet papr.
We follow a typical or default way to split these datasets into training and validation data sets with
standard data augmentation scheme which is widely used for these datasets~\cite{sermanet2012convolutional, he2016deep, he2016identity, huang2017densely}. 
%\paragraph{MNIST.}
%The MNIST datasetconsist of grayscale images with handwritten digits with 
%$28\times28$ pixels. There are 10 classes for these labels. The training and test set contain 60,000 and
%10,000 images. 
%\paragraph{CIFAR.}
%The two CIFAR datasets~\cite{krizhevsky2009learning} consist of colored natural images with $32\times32$ pixels. 
%CIFAR-10 consists of images drawn from 10 and CIFAR-100 from 100 classes. 
%The training and test sets contain 50,000 and 10,000 images respectively. 
%We adopt a standard data augmentation scheme that is widely used for these two datasets~\cite{sermanet2012convolutional, he2016deep, he2016identity, huang2017densely}. 
%For the final run we use all 50,000 training images and report the final test error at the end of training.

\paragraph{Models implementation}
%I suggest to put some comments about where to put BN and whether to use Dropout.
%Or some other tricks that mentioned like in "implementation detials" in DenseNet.
In our experiments, the structure of classical ResNet or pre-act ResNet models are implemented with
the same structure as in the sample codes in PyTorch or Torchvision. 
As for our modified models, we implement them after some
modifications of these standard codes.  
Following the strategy in~\cite{he2016deep, he2016identity}, we adopt Bath Normalization~\cite{ioffe2015batch} 
but no Dropout~\cite{srivastava2014dropout}.

\paragraph{Training details}
We adopt the SGD training algorithm with momentum of $0.9$ and the weight initialization strategy as in~\cite{he2015delving}. We also adopt weight decay, which are $0.0001$ for ResNet18 type models and $0.001$ for ResNet34 type models. We take  minibatch size to be 128, 128, 256 for  MNIST, CIFAR and ImageNet, respectively.
We start with a learning rate of 0.1, divide it by 10 every 30 epochs, and terminate training at 60 epochs for MNIST and 120 epochs for CIFAR. On ImageNet, we start with a learning rate of 0.1, divide it by 10 every 40 epochs, and terminate training at 120 epochs.

%For MNIST, these models are trained with minibatch size of 128. We start with a learning rate of 0.1, divide it by 10 at 30 epochs, and terminate training at 60 epochs. 
%For CIFAR10, we start with a learning rate of 0.1. For ResNet, the minibatch size is 128 and we divide learning rate by 10 at 30 epochs, and terminate training at 120 epochs.  
%For CIFAR100, these models are trained with a minibatch size of 128. We start with a learning rate of 0.1, divide it by 10 at 30 epochs, and terminate training at 120epochs.
%For DenseNet, the minibatch size is 64 and we divide learning rate by 10 at 150, 225 epochs, and terminate training at 300 epochs.

\subsection{Classification accuracy on dataset for modified models}
%I suggest to merge the numerical results for CIFAR-10 and CIFAR-100 together 
%like the Table 2 in DenseNet.
This modified pre-act ResNet can also be understood as a special
parameter sharing form on $A^{\ell,i}$. With the similar idea, we want to prove that
the linear model real makes sense not because of the redundancy of CNNs.
Thus, we also put this parameter sharing scheme to $B^{\ell,i}$ or both $A^{\ell,i}$ and $B^{\ell,i}$
for pre-act ResNet. 
\begin{description}
	\item[Pre-acc ResNet-$A^{\ell,i}$-$B^\ell$:]  \quad 
	%	\begin{equation}
	$r^{\ell,i} =r^{\ell, i-1} + A^{\ell,i} \ast \sigma \circ B^{\ell} \ast \sigma (r^{\ell,i-1}), \quad i = 1:\nu_\ell.$
	%	\end{equation}
	\item[Pre-acc ResNet-$A^\ell$-$B^\ell$:] \quad 
	%	\begin{equation}
	$r^{\ell,i} =r^{\ell, i-1} + A^{\ell} \ast \sigma \circ B^{\ell} \ast \sigma (r^{\ell,i-1}), \quad i = 1:\nu_\ell.$
	%	\end{equation}
\end{description}
The corresponding architectures for ResNet is defined in the same fashion. Because of the special 
role for $B^{\ell,0}$, we only share $B^{\ell,i}$ for $ i = 1:\nu_\ell$.
For simplicity and consistency, we denote the classical ResNet and pre-act ResNet as
ResNet-$A^{\ell,i}$-$B^{\ell,i}$ and pre-act ResNet-$A^{\ell,i}$-$B^{\ell,i}$.
\begin{table}[H]
	\caption{The TOP-1 accuracy and number of parameters of ResNet-18, pre-act ResNet-18 and their modified models on MNIST.}
	\label{MNIST}
	\begin{center}
		\begin{tabular}{lcc}
			\toprule
			Model & Accuracy & \# Parameters \\ 
			%			\hline\hline
			\midrule
			ResNet18-$A^{\ell,i}$-$B^{\ell,i}$      			& 99.56 	& 11M  \\ 
			%			\hline
			ResNet18-$A^\ell$-$B^{\ell,i}$ 			& {\bf 99.58} 	& 8.0M   \\ 
			\midrule
			pre-act ResNet18-$A^{\ell,i}$-$B^{\ell,i}$       		& 99.61 	& 11M  \\ 
			%			\hline
			pre-act ResNet18-$A^\ell$-$B^{\ell,i}$	& {\bf 99.64} 	& 8.0M   \\ 
			\bottomrule
			%			\hline
			%			DenseNet 				& 99.59 	& 0.11M   	\\ \hline
			%			DenseNet-$A^\ell$ 			& 99.59 	& 72K   	\\ \hline
		\end{tabular} 
	\end{center}
\end{table}
%\subsubsection{Cifar10}

\begin{table}[H]
	\caption{The TOP-1 accuracy and number of parameters for ResNet, pre-act ResNet and their variants of modified versions on CIFAR10 and CIFAR100.}
	\label{CIFAR}
	\begin{center}
		\begin{tabular}{lccc}
			\toprule
			Model   					&  CIFAR10 & CIFAR100 & \# Parameters\\ 
			%			\hline \hline
			\midrule
			ResNet18-$A^{\ell,i}$-$B^{\ell,i}$       		& 93.45 		& 74.45			& 11M   \\ 
			%			\hline
			ResNet18-$A^\ell$-$B^{\ell,i}$ 					& {\bf 93.54} & {\bf 74.46}  & 8.1M	\\ 
			%			\hline
			ResNet18-$A^{\ell,i}$-$B^\ell$ 					& 93.35 		& 72.78 		 & 9.7M	\\ 
			%			\hline
			ResNet18-$A^\ell$-$B^\ell$ 						& 93.32  		& 72.56			 & 6.6M \\ 
			\midrule
			pre-act ResNet18-$A^{\ell,i}$-$B^{\ell,i}$ & 93.75 			& 74.33			 & 11M	 \\ 
			%			\hline
			pre-act ResNet18-$A^\ell$-$B^{\ell,i}$ 	    & {\bf 93.83} & {\bf 74.51}	  & 8.1M  \\ 
			%			\hline
			pre-act ResNet18-$A^{\ell,i}$-$B^\ell$ 		& 93.80 	    & 72.67          & 9.7M	\\ 
			%			\hline
			pre-act ResNet18-$A^\ell$-$B^\ell$ 		    & 93.45 	    & 72.81          & 6.6M	\\ 
			\midrule
			ResNet34-$A^{\ell,i}$-$B^{\ell,i}$       		& 94.71 		& 77.20			 & 21M	\\ 
			%			\hline
			ResNet34-$A^\ell$-$B^{\ell,i}$ 					& {\bf 94.84} & {\bf 77.24}	 & 13M	\\ 
			%			\hline
			ResNet34-$A^{\ell,i}$-$B^\ell$ 					& 93.94 	    & 75.31		     & 15M	\\ 
			%			\hline
			ResNet34-$A^\ell$-$B^\ell$ 						& 93.79  	     & 74.79		 & 6.7M	\\ 
			\midrule
			pre-act ResNet34-$A^{\ell,i}$-$B^{\ell,i}$ & 94.76 			& 77.25			& 21M	\\ 
			%			\hline
			pre-act ResNet34-$A^\ell$-$B^{\ell,i}$ 		& {\bf 94.84} & {\bf 77.40}  &13M	\\ 
			%			\hline
			pre-act ResNet34-$A^{\ell,i}$-$B^\ell$		& 93.94		    & 75.36			& 15M	\\ 
			%			\hline
			pre-act ResNet34-$A^\ell$-$B^\ell$ 			& 93.75		    & 75.82			& 6.7M	\\ 
			\bottomrule
		\end{tabular} 
	\end{center}
\end{table}

\begin{table}[H]
	\caption{The TOP-1 accuracy and number of parameters for ResNet, pre-act ResNet and their variants of modified versions on ImageNet.}
	\label{ImageNet}
	\begin{center}
		\begin{tabular}{lcc}
			\toprule
			Model   					& Accuracy  & \# Parameters \\	
			\midrule
			ResNet18-$A^{\ell,i}$-$B^{\ell,i}$       	&   70.23	    &  11.7M  		\\ 
			%			\hline
			ResNet18-$A^{\ell}$-$B^{\ell,i}$			& 	69.45	    &  8.6M   		\\ 
			\midrule
			pre-act ResNet18-$A^{\ell,i}$-$B^{\ell,i}$  &   70.46		&  11.7M			\\ 
			%			\hline
			pre-act ResNet18-$A^{\ell}$-$B^{\ell,i}$    &  	69.78		&  	8.6M			\\ 
			\bottomrule
		\end{tabular}
	\end{center}
\end{table}

%\begin{table}[t]
%	\begin{center}
%		\begin{tabular}{|l|cc|}
%			\hline
%			Model   					& Accuracy  & \# Parameters  \\	\hline\hline
%			ResNet18-$A^{\ell,i}$-$B^{\ell,i}$       				& 74.45		& 11M  	\\ \hline
%			ResNet18-$A^\ell$-$B^{\ell,i}$ 				& {\bf 74.46} 	& 8.1M 	\\ \hline
%			ResNet18-$A^{\ell,i}$-$B^\ell$ 				& 72.78  	& 9.8M 	\\ \hline
%			ResNet18-$A^\ell$-$B^\ell$ 		& 72.56  	& 6.7M 	\\ \hline
%			pre-act ResNet18-$A^{\ell,i}$-$B^{\ell,i}$       			& 74.33 	& 11M  	\\ \hline
%			pre-act ResNet18-$A^\ell$-$B^{\ell,i}$ 			& {\bf 74.51} 	& 8.1M  \\ \hline
%			pre-act ResNet18-$A^{\ell,i}$-$B^\ell$ 			& 72.67 	& 9.8M  \\ \hline
%			pre-act ResNet18-$A^\ell$-$B^\ell$ 		& 72.81  	& 6.6M 	\\ \hline
%			ResNet34-$A^{\ell,i}$-$B^{\ell,i}$       				& 77.20 	& 21M \\ \hline
%			ResNet34-$A^\ell$-$B^{\ell,i}$ 				&{\bf  77.24} 	& 13M \\ \hline
%			ResNet34-$A^{\ell,i}$-$B^\ell$ 				& 75.31		& 15M \\ \hline
%			ResNet34-$A^\ell$-$B^\ell$ 		& 74.79   	& 6.7M \\ \hline
%			pre-act ResNet34-$A^{\ell,i}$-$B^{\ell,i}$       			& 77.25 	& 21M \\ \hline
%			pre-act ResNet34-$A^\ell$-$B^{\ell,i}$ 			& {\bf 77.40} 	& 13M   \\ \hline
%			pre-act ResNet34-$A^{\ell,i}$-$B^\ell$ 			& 75.36 	& 15M  \\ \hline
%			pre-act ResNet34-$A^\ell$-$B^\ell$ 		& 75.82 	& 6.7M   \\ \hline
%		\end{tabular}
%	\end{center}
%	\caption{The accuracy and number of parameters for models on CIFAR100.}
%\end{table}
From these numerical results, we have the next two important observations: (1) The modified ResNet and pre-act ResNet models achieve almost the same accuracy to the original models on different datasets as shown in Tables \ref{MNIST}, \ref{CIFAR} and \ref{ImageNet}. (2) Only models with $A^{\ell}$ type can keep the accuracy, which is consistent with our analysis of the constrained linear data-mapping. Any models formally modified by changing $B^{\ell,i}$ to $B^\ell$ have lower accuracy, especially on CIFAR100. These observations indicate that constrained data-feature mapping together with the relevant iterative methods can provide mathematical insights of the ResNet and pre-act ResNet models.

\section{Discussion and Conclusion}
In this paper, we propose a constrained linear data-feature mapping model 
behind CNN models for image classification such as ResNet. 
Under this assumption, we carefully study the connections between the traditional
iterative method with nonlinear constraint and the basic block scheme in pre-act ResNet model, 
and make an explanation for pre-act ResNet in a technical level. 
Comparing with other existing works that discuss the
connection between dynamic systems and ResNet, the constrained data-feature
mapping model goes beyond formal or
qualitative comparisons and identifies key model components with much more details.  
Furthermore, we hope that how and why ResNet type models work can be
mathematically understood in a similar fashion as for classical 
iterative methods in scientific computing which has a much more 
mature and better developed theory.  
Some numerical experiments are verified in this paper which indicate the rational and efficiency
for the constrained learning data-feature mapping model. 

We hope our attempt about the connection of CNNs and classical iterative methods
can open a new door to the mathematical understanding, analysis and improvements of 
CNNs with some special structures.  These results presented in
this paper have demonstrated the great potential of this model from both
theoretical and empirical viewpoints.  
Obviously many aspects of classical iterative methods with constraint
should be further explored and expect to be much improved. 
For example, we are trying to establish the connection of DenseNet and
the the so-called multi-step iterative methods in numerical linear algebra~\cite{hackbusch1994iterative, golub2012matrix}.

\vspace{30pt}
\hspace{-1.5em}{\LARGE \bf{Appendix}}

\section*{A\quad Comparison between pooling block of ResNet and restriction in multigrid}
\label{sec:pooling}
So far, we investigate the basic iterative block for pre-act ResNet
from the data-feature mapping perspective. We now try to involve the
pooling block into this framework by introducing the multi-scale structure
as in multigrid~\cite{xu1992iterative,hackbusch2013multi}.
We will now make comparison between the pooling block in pre-act ResNet 
with the standard restriction in multigrid for residual.
Here, we first introduce the pooling block in modified pre-act ResNet as
\begin{equation}\label{pooling-preact}
\tilde r^{\ell+1,0} =  \tilde R_\ell^{\ell+1} \ast_2  r^{\ell, \nu_\ell} + \tilde A^{\ell+1} \ast \sigma \circ \tilde B^{\ell+1,0}\ast _2 \sigma ( r^{\ell, \nu_\ell} ).
\end{equation}
While, using the pooling of residual in multigrid, we have
\begin{equation}
r^{\ell+1,0} = R_\ell^{\ell+1} \ast_2 (f^\ell - A^\ell(u^{\ell, \nu_\ell})) = R_\ell^{\ell+1} \ast_2 r^{\ell,\nu_\ell}.
\end{equation}
Take this into feature extraction, we have
\begin{equation}
r^{\ell+1,1} = r^{\ell+1,0} + A^{\ell+1}\ast \sigma \circ B^{\ell+1,1} \ast \sigma (r^{\ell+1,0}).
\end{equation}
This means that
\begin{equation}
r^{\ell+1,1} = R_\ell^{\ell+1} \ast_2 r^{\ell,\nu_\ell} + A^{\ell+1}\circ \sigma \circ B^{\ell+1,1} \ast \sigma (R_\ell^{\ell+1} \ast_2 r^{\ell,\nu_\ell}).
\end{equation}
By taking $\tilde R_\ell^{\ell+1} = R_\ell^{\ell+1}$ and $\tilde A^{\ell+1} =  A^{\ell+1}$, we have
\begin{equation}\label{pooling-preact1}
\tilde r^{\ell+1,0} =  R_\ell^{\ell+1} \ast_2 r^{\ell, \nu_\ell} +  A^{\ell+1} \ast \sigma \circ \tilde B^{\ell+1,0}\ast_2  \sigma ( r^{\ell, \nu_\ell} ).
\end{equation}
Thus, the difference between $\tilde r^{\ell+1,0}$ and $r^{\ell+1,1}$ is noted in 
the difference of 
\begin{equation}\label{eq:diff1}
\tilde B^{\ell+1,0}\ast_2  \sigma ( r^{\ell, \nu_\ell} ) ~~\text{and}~~B^{\ell+1,1} \ast \sigma (R_\ell^{\ell+1} \ast_2 r^{\ell,\nu_\ell}).
\end{equation}
Here let us ignore the nonlinear activation and rewrite the convolution with stride $2$ as
\begin{equation}\label{key}
\tilde B^{\ell+1,0}\ast_2  r^{\ell, \nu_\ell} = S(\tilde B^{\ell+1,0}\ast  r^{\ell, \nu_\ell} ),
\end{equation}
where $S$ means sub-sampling such as $[S(r)]_{i,j} = r_{2i-1, 2j-1}$.
Thus the difference in \eqref{eq:diff1} becomes
\begin{equation}
S(\tilde B^{\ell+1,0}\ast  r^{\ell, \nu_\ell}) ~\text{and}~B^{\ell+1,1} \ast  R_\ell^{\ell+1} \ast S(r^{\ell,\nu_\ell}),
\end{equation}
because of the fact that $R_\ell^{\ell+1}$ chooses $1\times1$ kernel in pre-act ResNet. 
Let use consider that $r^{\ell,\nu_\ell} \in \mathbb{R}^{n_\ell\times n_\ell \times1}$ and 
$[r^{\ell,\nu_\ell} ]_{i,j} = 0$ expect for $[r^{\ell,\nu_\ell} ]_{2,2} = 1$, 
then $S(r^{\ell,\nu_\ell}) = {\bm 0} \in \mathbb{R}^{\frac{n_\ell}{2} \times \frac{n_\ell}{2} \times1}$.
However, we may learn some special $\tilde B^{\ell+1,0}$ such that 
$S(\tilde B^{\ell+1,0}\ast  r^{\ell, \nu_\ell}) \neq \bm 0$ which can capture the one pixel feature.
From this point of view, we may say that the pooling block \eqref{pooling-preact} in pre-act ResNet 
makes sense to prevent loosing of small scale information.
Thus, we will choose the pooling block in  \eqref{pooling-preact} to be
the pooling block of modified ResNet ,pre-act ResNet  or other
models without special statements.

\bibliographystyle{abbrv}
\bibliography{LinearModel_Jul06}

\end{document}